\documentstyle[preprint,aps]{revtex}
\draft
\begin{document}
\title{The Broken Symmetry of Two-Component $\nu=1/2$ Quantum Hall States}
\author{Tin-Lun Ho}
\address{Physics Department,  The Ohio State University, Columbus, Ohio
43210}
\maketitle

\begin{abstract}
We show that the recently discovered $\nu=1/2$
quantum Hall states in bilayer systems
are triplet p-wave pairing states of composite Fermions,
of exactly the same form as $^{3}$He superfluids.
The observed persistence (though weakening) of the $\nu=1/2$ state
in  the two- to one-component crossover region corresponds to a continuous
deformation of the so-called (331) state towards the ``Pfaffian" state,
identical to the well known A to A$_{1}$ transition  in $^{3}$He.
This deformation also demonstrates the remarkable fact that
electrons can release and capture
``vortices" in a continuous and incompressible manner through
spin rotations.
The broken symmetry of the triplet pairing state is
a ``pairing" vector ${\bf d}$. It also implies a (pseudo-spin) magnetization
$\propto i{\bf d}\times{\bf d}^{\ast}$.  In the presence of layer tunneling,
the (331) state ({\bf d} real) is unstable against other states with a
magnetization ({\bf d} complex).
The recently observed persistence of the $\nu=5/2$ state in single layer
systems
in the two- to one-component crossover region
is also consistent with triplet pairing interpretation.
\end{abstract}

\vspace{0.2in}


\bigskip

\pagebreak

The robust $\nu=1/2$ quantum Hall state  recently discovered in bilayer
quantum well systems\cite{ATT} and the $\nu=5/2$ state
discovered\cite{Willett} some years ago in
single layer systems have illustrated the dramatic effect of
internal degrees of freedom on the quantum Hall phenomenon, namely,
the formation of multicomponent quantum Hall states.
The 5/2 state was discovered in single-layer systems with weak Zeeman
energy, where both up and down spins interact strongly with each other and
are believed to play equal role in the makeup of the 5/2 state.
For bilayer systems, even though the electron spins are frozen by the
magnetic field, they behave effectively as  pseudo-spin
1/2 systems. The two pseudo-spin states correspond to
the symmetric and antisymmetric state of the quantum well.
The $\nu=1/2$ state was first discovered in the
``two-component" regime\cite{ATT} where the pseudo-spins have a  weak
Zeeman energy.  Because of the similarity of these two systems, we shall
in the following refer both pseudo-spins and real electron spins simply as
``spins".

These two systems, however,
are characterised by different energy parameters.
In the absence of disorder, bilayer systems have three
energy scales: (a) the energy difference $\Delta$ between the symmetric
and antisymmetric states,
(which is the ``tunneling energy" between the layers);
(b) intra-layer Coulomb interaction $e^{2}/\ell$, where $\ell$ is the
magnetic length; and (c) inter-layer interaction $e^{2}/D$, where $D$ is
the separation between the layers. These scales in turn lead to three
distinct physical regimes :

\noindent {\bf \{1\}}+{\bf \{1\}} :
 Two separate one-component regime, $e^{2}/\ell>>e^{2}/D,\Delta$,

\noindent {\bf \{2\}} : Two-component regime,
$e^{2}/\ell\sim e^{2}/D >>\Delta$.

\noindent {\bf \{1\}} :
One single-component regime, $e^{2}/\ell, e^{2}/D < \Delta$.

\noindent
Regime ({\bf \{1\}}+{\bf \{1\}}) is realized for large $D$,
where the system separates into two
weakly coupled single-layer systems. Regime
{\bf \{2\}} is the case where intra- and inter-layer interactions are
comparable and dominate over the tunneling energy so that the two
spin-populations appear nearly degenerate.
Regime {\bf \{3\}} is the strong tunneling limit where all electrons lie in
the symmetric state. As mentioned before, the
$\nu=1/2$ bilayer state\cite{ATT} was first discovered
in regime {\bf \{2\}}. However, recent experiments\cite{Prin2} show that
even though the $\nu=1/2$ state is
weakened and eventually disappears as the system is tuned from regime
{\bf \{2\}}
to regime {\bf \{1\}} it remains a good quantum Hall
state (i.e. incompressible) for a sizable range of parameters
$\Delta/(e^{2}/\ell)$.
So far, theoretical studies of the bilayer $\nu=1/2$ state have
focused mainly on
energetics. There are little studies of  properties.
Numerical studies\cite{SongHe}
 indicate that the observed $\nu=1/2$ state is likely to be the
so-called (331) state\cite{Halperin}. There are also suggestions\cite{Prin2}
 that the observed persistence of the
 $\nu=1/2$ state in the
 {\bf \{2\}} to {\bf \{1\}} crossover region correspond to the evolution of
 the (331) state into the so-called Pfaffian state\cite{ReadMoore}.
No further analysis, however, has been made with this interpretation.
Moreover, this picture seems to run counter of the conventional view that
the (331) state can not evolve continuously into the Pfaffian state as they
have different  ``topological order"\cite{Wen}.

The purpose of this paper to point out a fundamental broken symmetry
 of the bilayer $\nu=1/2$ states and  the important properties
associated with it. This symmetry is not obvious in the conventional
$(331)$ and Pfaffian representation because
they  make reference to a specific spin quantization axis. To reveal this
symmetry, it is necessary to use a general representation independent
of spin quantization axes. The broken symmetry revealed leads to the following
results :

\noindent ({\bf I})  Both (331) and Pfaffian state are triplet p-wave  BCS
states of composite Fermions\cite{Jain}
with wavefunctions identical to that of the A
and the A$_{1}$ phase of superfluid $^{3}$He\cite{Voll}.
The transformation from the
(331) to the Pfaffian state corresponds to the well known A to A$_{1}$
transition in $^{3}$He. The broken symmetry of triplet BCS superfluids is
a complex vector ${\bf d}$ (referred to as ``pairing vector") denoting
the direction of zero spin projection of the pair,
i.e. $|S=1, {\bf d}\cdot {\bf S}=0>$, where ${\bf S}$ is the total spin of the
pair. The ${\bf d}$ vector of the (331) and the Pfaffian state is
$\hat{\bf x}_{3}$ and
$(\hat{\bf x}_{3}-i\hat{\bf x}_{2})/\sqrt{2}$ respectively.

\noindent ({\bf II})
Contrary to the current topological order arguments\cite{Wen},
which implies that the (331) state can not  be deformed continuously
into the Pfaffian state because they have different degeneracies on a torus,
we  show (from the explicit solution of a model
Hamiltonian which reflects the essential physics of the bilayer system)
that $continuous$ transition between (331) and Pfaffian can indeed occur.
In other words,
there are neither topological  nor intrinsic energetic obstructions
to prevent continuous transformation between any two triplet pairing states.

[Our explicit proof shows that
degeneracies in toroidal geometries does not
truly reflect the topological properties of the quantum Hall states with
internal degrees of freedom.
The different degeneracies at different points of the triplet pairing space
can  reconcile with each other inside the triplet space.
The situation is similar to the topological stability of the vortices in
x-y spin systems. Vortices with different winding numbers are topological
distinct if the spins are confined in the
x-y plane. However, they can be deformed into each other if
 the spins are given a z-component to become Heisenberg spins].

\noindent ({\bf III}) In the presence of tunneling, (however weak),
the (331) state
(with a real ${\bf d}$) is unstable with respect to the appearance of
an imaginary component in  ${\bf d}$. The
implication of a complex ${\bf d}$ is that the system will  have a
magnetization ${\bf m}$ proportional to $i{\bf d}\times{\bf d}^{\ast}$,
which grows from zero to a saturating value equal to half the full density
$(\overline{\rho}/2=\nu/4\pi=1/8\pi)$ as the (331) state evolves towards
the Pfaffian state.
Thus, in the presence of tunneling, the $1/2$ state will acquire a
magnetization like that of the $\nu=1$ state\cite{ATT}\cite{ATT2}.
As this magnetization grows, the $\nu=1/2$ state should exhibit properties
typical of quantum Hall ferromagnets\cite{WenZee}\cite{Yang}\cite{Ho}.

We shall also discuss the implication of
triplet pairing on the single layer 5/2 state\cite{Willett} at the end of this
paper.

To begin, let us first review the dynamics of the bilayer systems in the
pseudo-spin language\cite{cHo}. In the
lowest Landau level, the Hamiltonian of a system of $2N$ electrons
is a sum of the interaction
energy $V$ and the tunneling energy $H_{T}$, where
\begin{equation}
V = \sum_{i>j=1}^{2N} \left[
V_{o}({\bf R}_{i}-{\bf R}_{j}) + \sum_{\alpha=1}^{3}
V_{\alpha}({\bf R}_{i}-{\bf R}_{j})S_{\alpha}(i)S_{\alpha}(j)\right],
\,\,\,\,\,\, H_{T}= - \Delta \sum_{i=1}^{2N}S_{1}(i)
\label{pot}
\end{equation}
where ${\bf R}_{i}$ is the guiding center coordinate of the $i$th electron.
We have chosen a coordinate system $\hat{\bf x}_{1}, \hat{\bf x}_{2},
\hat{\bf x}_{3}$ such that the symmetric and antisymmetric states are
represented by  the spinors $(^{1}_{1})$ and $(^{1}_{-1})$ respectively, and
that the layer tunneling acts like a magnetic field along $\hat{\bf
x}_{1}$.
The states localized in the upper and lower layer will then correspond to
$(^{1}_{0})$ and $(^{0}_{1})$, and
$S^{\rm total}_{3}\equiv\sum_{i}S_{3}(i)=(N_{\uparrow}-N_{\downarrow})/2$
 is the difference in electron number between the two layers.
For typical parameters of the quantum wells
(layer separation $D$, well depth, etc), $V_{3}>>|V_{1}|, |V_{2}|$.
The Coulomb interaction  $V_{3}$ favors equal
electron densities on both layers, (ie.
 $<S^{\rm total}_{3}>=0$). On the other hand,
$\Delta$ favors a maximum $<S^{\rm total}_{1}>$.

In regime {\bf \{2\}},  numerical studies\cite{SongHe} indicate that
the $\nu$=1/2 state is the (331) state,
\begin{equation}
|\Psi_{331}> = \int
 \prod_{i>j=1}^{N}(a_{i}-a_{j})^{3}(b_{i}-b_{j})^{3}
\prod_{i,j=1}^{N}(a_{i}-b_{j})^{1}
\prod_{i=1}^{N}\psi^{+}_{\uparrow}(a_{i})\psi^{+}_{\downarrow}(b_{i})
|0> \label{331} ,
\end{equation}
where $\int \equiv \int \prod_{i=1}^{N}\left(
d^{2}a_{i}d^{2}b_{i}\right)
{\rm exp}(-\frac{1}{4}\sum^{N}_{i=1}(|a_{i}|^{2}+|b_{i}|^{2}))$ represents
the integration over all electrons with their Gaussian factors included.
In regime {\bf \{1\}},
it is expected the $\nu$=1/2 state will be close to the Pfaffian form,
which is a spin polarized state
\begin{equation}
|\Psi_{\rm Pf}> = \int \Phi^{2}{\rm Pf}\left(\frac{1}{z_{i}-z_{j}}\right)
\prod_{i=1}^{2N}\psi^{+}(z_{i})|0>
\label{Pf} \end{equation}
where
$\psi^{+}(z)=[\psi^{+}_{\uparrow}(z)+\psi^{+}_{\downarrow}(z)]/\sqrt{2}$
represents the symmetric state of the quantum well,  and
$\Phi\equiv\prod_{i>j=1}^{2N}(z_{i}-z_{j})$ is the Laughlin factor
for all electrons. The Pfaffian
of a matrix $M_{i,j}$ is defined as ${\rm Pf}(M_{i,j})
=\sum_{P}\prod^{N}_{k=1}M_{P(2k-1), P(2k)}$,  where $P$ is a permutation of
$2N$ objects.

We now show that both eq.(\ref{331}) and eq.(\ref{Pf}) belong to the same
family of triplet pairing states,
\begin{equation}
|N,{\bf d}>\equiv|N, \chi> =\int \Phi^{2} \prod_{i=1}^{N}
\left[\frac{1}{(z_{2i-1}-z_{2i})}\chi^{}_{\mu\nu}
\psi^{+}_{\mu}(z_{2i-1})\psi^{+}_{\nu}(z_{2i})\right] |0> ,
\label{tri}
\end{equation}
where $\chi$ is a symmetric $2\times 2$ matrix (which can be normalized as
Tr$\chi^{+}\chi$=2 without loss of generality). Since the general
representation of a 2$\times$2 symmetric matrix is
\begin{equation}
\chi_{\mu\nu}= i{\bf d}\cdot[\vec{\sigma}\sigma_{2}]_{\mu\nu},
\label{chi}
\end{equation}
where ${\bf d}$ is a complex vector,
the triplet space is simply
the space of the ``pairing"  vector $\{ {\bf d}\}$, ($|{\bf d}|^{2}=1$).
{}From the property
$<\chi|{\bf S}|\chi>/<\chi|\chi>$=
$\frac{1}{2}{\rm Tr}[\chi^{+}({\bf S}(1)+{\bf S}(2))\chi]
=2i{\bf d}\times{\bf d}^{\ast}$, one can see that ${\bf d}$ represents the
the direction of zero spin projection, and that average spin
is nonzero if and only if ${\bf d}$ is complex.

To see the relation between eq.(\ref{Pf}) and eq.(\ref{tri}), we note that
each term in the Pfaffian contributes identically in eq.(\ref{Pf}).
Hence,  $|\Psi_{\rm Pf}>=$
 $(2N!)$$\int\Phi^{2}$$M_{12}M_{34}..M_{2N-1, 2N}$
$[\psi^{+}(z_{1})\psi^{+}(z_{2})][\psi^{+}(z_{3})\psi^{+}(z_{4})]..
[\psi^{+}(z_{2N-1})\psi^{+}(z_{2N})]|0>$,  where $M_{ij}=(z_{i}-z_{j})^{-1}$.
Apart from a normalization constant, this is precisely eq.(\ref{tri}) with
$\chi=2^{-1/2}(^{1}_{1}$$^{1}_{1})$, or
${\bf d}=(\hat{\bf x}_{3}-i\hat{\bf x}_{2})/\sqrt{2}$.
To see the relation between eq.(\ref{331}) and (\ref{tri}), we make use of
the Cauchy identity\cite{Muir}
\begin{equation}
\prod_{i>j=1}^{N}[(a_{i}-a_{j})(b_{i}-b_{j})]
=\prod_{i,j=1}^{N}(a_{i}-b_{j}){\rm Det}|(a_{i}-b_{j})^{-1}|. \label{Cauchy}
\end{equation}
This identity states that the vortices of electrons of like
spins can turn into those of unlike spins provided the latter
pair up as described by the determinant. Using
eq.(\ref{Cauchy}), the integrand of eq.(\ref{331}) can  be written as
$\Phi^{2}{\rm Det}|M_{ij}|$. Proceeding similar to the Pfaffian case, we can
write
$|\Psi_{331}>=$$(N!)\int\Phi^{2}M_{12}M_{32}..M_{2N-1,2N}$
$[\psi^{+}_{\uparrow}(1)\psi^{+}_{\downarrow}(2)]$
$[\psi^{+}_{\uparrow}(3)\psi^{+}_{\downarrow}(4)]..$
$[\psi^{+}_{\uparrow}(2N-1)\psi^{+}_{\downarrow}(2N)]|0>$, where
$\psi^{+}_{\mu}(i)=\psi^{+}_{\mu}(z_{i})$.
Because $M_{ij}$ is antisymmetric, the product
$\psi^{+}_{\uparrow}(1)\psi^{+}_{\downarrow}(2)$ in the integral
can be replaced by the triplet state
$\frac{1}{2}[\psi^{+}_{\uparrow}(1)\psi^{+}_{\downarrow}(2)
+\psi^{+}_{\downarrow}(1)\psi^{+}_{\uparrow}(2)]$, which takes
$|\Psi_{331}>$ into the pairing form
 eq.(\ref{tri}), with $\chi=(^{0}_{1}$$^{1}_{0})$, i.e.  ${\bf d}=
\hat{\bf x}_{3}$. Thus, we have
\begin{equation}
|\Psi_{331}> = |N,{\bf d}=\hat{\bf x}_{3}>, \,\,\,\,\,\,\,\,\,\,
[\Psi_{\rm Pf}>=|N,{\bf d}=\frac{\hat{\bf x}_{3}-i\hat{\bf x}_{2}}{\sqrt{2}}>.
\label{result} \end{equation}

The triplet pairing family (eq.(\ref{tri})) can be written in a more
concise form in terms of
composite Fermion operator $\phi^{+}_{\mu}(z) = \psi^{+}_{\mu}(z)U(z)^{2}$,
where $U(z)$ is the quasihole operator\cite{ReadGL}  defined as
\begin{equation}
U(a)\psi^{+}_{\mu}(z)=(a-z)\psi^{+}_{\mu}(z)U(z), \,\,\,\,\,\,
\psi_{\mu}(z)U(a)=(a-z)U(a)\psi_{\mu}(z), \,\,\,\,\,\, U(a)|0>=|0> .
\label{qh} \end{equation}
In terms of $\phi^{+}_{\mu}$, eq.(\ref{tri}) reduces to
\begin{equation}
|N, \chi> = |N,{\bf d}> = (Q^{+})^{N}|0> ,
\label{he3}
\end{equation}
\begin{equation}
Q^{+} = \int \chi^{}_{\mu\nu}(z,z')\phi^{+}_{\mu}(z)
\phi^{+}_{\nu}(z'), \,\,\,\,\,\,\, \chi^{}_{\mu\nu}(z,z')=
(z-z')^{-1} \chi^{}_{\mu\nu} ,
\label{Q} \end{equation}
where $\int\equiv\int d^{2}z d^{2}z' e^{-(|z|^{2}+|z'|^{2})/4}$.
Eq.(\ref{he3})  represents a triplet p-wave
BCS state of composite Fermions. If the composite Fermions
in eq.(\ref{he3}) were ordinary Fermions,
$|\Psi_{331}>$ and $|\Psi_{\rm Pf}>$ reduce to the ground states of
of $^{3}$He-A and $^{3}$He-A$_{1}$ in a thin slab. The slab geometry fixes
the orbital angular momentum axis
of the Cooper pairs of $^{3}$He-A and $^{3}$He-A$_{1}$
so that their orbital wavefunctions are given by
$x-iy$. The geometry has no effect on the pairing vector, which is
$\hat{\bf x}_{3}$ for $^{3}$He-A and ($\hat{\bf x}_{3}
-i\hat{\bf x}_{2})/\sqrt{2}$ for $^{3}$He-A$_{1}$\cite{Voll}).

Note that the entire triplet pairing family eq.(\ref{tri}) has filling factor
$\nu=1/2$, because of the factor $\Phi^{2}$.
Since the triplet pairing space $\{ {\bf d}\}$ is simply connected, any two
states in this space can be deformed continuously
into one another in an incompressible
manner unless  prevented by energetic reasons.
In other words, in the absence of energy obstructions, quantum phase
transitions between two triplet pairing states are either non-existent or at
most of second (but not first) order.
{}From eq.(\ref{331}) and eq.(\ref{Pf}), one can see that
the relative angular momentum $L$
between two electrons of like spins is at least three
in the (331) state and whereas it starts with  $L=1$ between two
electrons in the Pfaffian
state. The deformability between these two states within the triplet
pairing space illustrates clearly the fact that {\em electrons can
release and capture relative angular momentum in a continuous and
incompressible manner through spin rotations}.

An intrinsic ``energy
obstructions", however, is nonexistent. To show this, we
construct below a $continuous$ family of model
Hamiltonians $\{ \tilde{H}({\bf d})\}$ with
 $\{ |N,{\bf d}>\}$ as their ground states. (One has in mind that the
effect of changing the external parameters is to change the vector ${\bf
d}$ in the Hamiltonian).
The continuity of $\tilde{H}({\bf d})$
then implies that the system can evolve from one
${\bf d}$-state to another continuously without changing its filling factor
while remaining in the ground state during the entire process.

To motivate this model Hamiltonian,
it is useful to look at the general structure
of the pair potential $\hat{V}(ij)$. Decomposing it into  relative
angular momentum channel ($L$), and diagonalize it in spin space within
each $L$ channel, $\hat{V}(ij)$ can be expressed in  the diagonal form
\begin{equation}
\hat{V}(ij) = \sum_{L=0,2,4...}P_{L}(ij) V_{L}^{s}
|S=0><S=0| +
\sum_{L=1,3,5,..}P_{L}(ij) \sum_{\alpha=1}^{3} V^{t}_{L,\alpha}
|{\bf d}_{L,\alpha}><{\bf d}_{L,\alpha}|
\end{equation}
where $P_{L}(ij)$ is the projection operator for relative angular momentum
$L$, and $\{ |{\bf d}_{L,\alpha}>, \alpha=1,2,3\}$ are the three triplet
eigenstates in channel $L$.
(The spin wavefunction of the singlet is
$(\sigma_{2})_{\mu\nu}=<\mu\nu|S=0>$. The wavefunctions of the triplet are
$(\chi^{(\alpha)}_{L})_{\mu\nu}=<\mu\nu|{\bf d}_{L,\alpha}>=
i{\bf d}_{L,\alpha}\cdot[\vec{\sigma}\sigma_{2}]_{\mu\nu}$. Orthogonality
between different triplet states implies
${\bf d}_{L,\alpha}\cdot{\bf d}_{L,\beta}^{\ast}=\delta_{\alpha\beta}$).
$V^{s}_{L}, V^{t}_{L,\alpha}$ are the pseudo-potentials for the singlet and
the triplet in channel  $L$. Like the triplet states
$\{ |{\bf d}_{L,\alpha}>\}$, they depend on the external parameters.

Consider now a model potential  $\hat{W}(ij)$ which consists only of  $s$ and
$p$ channels. All pseudo-potentials are positive definite except for
$V^{t}_{1}$, which is zero. Our model  Hamiltonian is then
$\tilde{H}({\bf d}_{1})=\sum_{i>j}\hat{W(ij)}$,
\begin{equation}
\hat{W}(ij) = V^{s}P_{0}(ij)|{\rm S=0}><{\rm S=0}|
+\sum_{\alpha=2,3}V^{t}_{\alpha}P_{1}(ij)|{\bf d}_{\alpha}><{\bf d}_{\alpha}|
,
\label{model} \end{equation}
where $V^{s}, V^{t}_{2}, V^{t}_{3}>0$. We now show that $|N,{\bf d}_{1}>$
is the ground state of eq.(\ref{model}). Since
$\tilde{H}({\bf d}_{1})$ is positive,
it is sufficient to show that $|N,{\bf d}_{1}>$ is annihilated by the
interaction of any particular pair, say, $\hat{W}(12)$.
 Defining $\chi^{(\alpha)}=<\mu\nu|{\bf d}_{\alpha}>$, the
wavefunction of $|N,{\bf d}_{1}>$ is
\begin{eqnarray}
\Psi\left(^{z_{1}}_{\mu_{1}}, ... , ^{z_{2N}}_{\mu_{2N}}\right)=
\Phi^{2} [
\chi^{(1)}(12)\chi^{(1)}(34)...\chi^{(1)}((2N-1)(2N)) \nonumber
\\
- \chi^{(1)}(13)\chi^{(1)}(24)...\chi^{(1)}((2N-1)(2N))
+ etc. ]
\label{wf} \end{eqnarray}
where $\chi^{(1)}(ij)=(z_{i}-z_{j})^{-1}\chi^{(1)}_{\mu_{i}\mu_{j}}$.
There are two kinds of terms in eq.(\ref{wf}). Those contain
$\chi^{(1)}(12)$ and those do not. The former will be annihilated by
$\hat{W}(12)$ as it is orthogonal to
$\hat{W}(12)$ in spin space. The latter can be expressed as
linear combination of the singlet
and the triplet states of 1 and 2,
(i.e. $(\sigma_{2})_{\mu_{1}\mu_{2}}$
and $\chi^{(\alpha)}_{\mu_{1}\mu_{2}}, \alpha=1,2,3$). However, all these
terms contain factors like  $(z_{1}-z_{2})^{p}(z_{1}+z_{2})^{n}$
with $p\geq 2$. They are therefore  annihilated by the relative angular
momentum projections operators in $\hat{W}(12)$.
We have thus shown that $|N,{\bf d}_{1}>$ is the ground state of
$\tilde{H}({\bf d}_{1})$, and have by now  established {\bf I} and {\bf II}
mentioned in the opening.

For bilayer systems, the actual path of evolution
from (331) to Pfaffian depends on how the external
parameters are varied. To determine the general feature of
this path, as well as to understand
other properties of the triplet pairing states, we note an important property
of the broken symmetry ${\bf d} -- $ that it gives rise to a magnetization
\begin{equation}
{\bf m} = \frac{1}{A}\frac{<N,{\bf d}|{\bf S}^{\rm total}|N,{\bf d}>}
{<N,{\bf d}|N,{\bf d}>}
 =  \lambda
\left(i{\bf d}\times{\bf d}^{\ast}\right)  \,\, ,
 \,\,\,\,\,\,\,\, {\bf S}^{\rm total}=\sum_{i=1}^{2N}{\bf S}(i) .
\label{mag}
\end{equation}
$A$ is the surface area. $\lambda$ is a function of $|{\bf d}^{2}|^{2}$
and  is of the order of $\overline{\rho}/2$,
where $\overline{\rho}=\nu/2\pi= 1/4\pi$ is the total electron density of the
two layers. (Eq.(\ref{mag}) can be shown in a straightforward manner using
eq.(\ref{tri})  and noting that the numerator in eq.(\ref{mag}) is of the form
${\rm Tr}(\vec{\sigma}(\chi\chi^{+})^{n}){\rm
Tr}(\chi\chi^{+})^{m}$).
Analytic evaluation of $\lambda({\bf d}^{2})$ in the thermodynamic limit
turns out to  be difficult. However, since the Pfaffian state (which has
${\bf d}^{2}=0$) is completely
spin polarized, $\lambda$ must satisfy $\lambda(0)=\overline{\rho}/2$.
Using symbolic computation, we have found
that for a four electrons system,
$\lambda = (\overline{\rho}/2)(1+\frac{49}{260}|{\bf
d}^{2}|^{2})^{-1}$.

In the same way of deriving eq.(\ref{mag}), one can show that the energy
(per unit area) of the system is
\begin{eqnarray}
E({\bf d})  = & f - g_{0}|{\bf d}\cdot\hat{\bf x}_{3}|^{2}
- g_{1}( {\bf d}^{2}({\bf d}^{\ast}\cdot\hat{\bf x}_{3})^{2} + c.c. )
\nonumber \\
 & + g_{2}|{\bf d}\times{\bf d}^{\ast}\cdot\hat{\bf x}_{3}|^{2}
- \Delta\lambda i{\bf d}\times{\bf d}^{\ast}\cdot\hat{\bf x}_{1} ,
\label{energy}
\end{eqnarray}
where $f, g_{0}, g_{1}, g_{2}$ and $\lambda$ are  functions of
$|{\bf d}^{2}|^{2}$.
Terms like $({\bf d}\cdot\hat{\bf x}_{3})(i{\bf d}\times{\bf d}^{\ast}\cdot
\hat{\bf x}_{3})$ are absent, (which is related to gauge invariance).
It is easy to see that the expectation value of the $V_{o}$ term in
eq.(\ref{pot}) with respect to $|N, {\bf d}>$
is only a function of ${\bf d}^{2}$, corresponding to the
$f$ term in eq.(\ref{energy}). The expectation of the $V_{1}$ term in
eq.(\ref{pot}) consists of two types of terms :
${\rm Tr}[(\chi\chi^{+})^{n}\sigma_{3}]{\rm Tr}[(\chi\chi^{+})^{m}\sigma_{3}]
+ (m \leftrightarrow n)$, and
${\rm Tr}[\chi(\chi^{+}\chi)^{n}\sigma_{3}(\chi^{+}\chi)^{m}\chi\sigma_{3}]
+(m \leftrightarrow n)$. The former give arise to the $g_{2}$ term in
eq.(\ref{energy}), and the latter the $g_{0}$ and $g_{1}$ terms.
On physical grounds, one can conclude that $g_{2} >0$. For otherwise the
 magnetization of the system will have a nonzero projection
along $\hat{\bf x}_{3}$, implying that the two layers have different electron
densities in equilibrium.
The fact that the system is in a (331) state in the
absence of layer tunneling\cite{SongHe} also means that the functional forms of
$f$ and $g$'s are such that $E({\bf d})$
 has a minimum at ${\bf d}=\hat{\bf x}_{3}$.

An immediate implication of eq.(\ref{mag}) and eq.(\ref{energy}) is  that
the (331) state is unstable against the appearance of a small imaginary
component in ${\bf d}$, i.e.  ${\bf d}=\hat{\bf x}_{3} \rightarrow
\hat{\bf x}_{3}-i\epsilon \hat{\bf x}_{2}, \,\,\, \epsilon<<1$.
This is because the system will then  acquire a magnetization along
$\hat{\bf x}_{1}$, thereby lowering its energy by an amount of order $\epsilon$
through the layer tunneling term $\Delta$ in eq.(\ref{energy}).
In contrast, the energy cost
in the Coulomb energy terms $V_{o}$ and $V_{3}$, (the $f$ and $g$ terms in
eq.(\ref{energy}))  are quadratic in
$\epsilon$. Thus, in the
presence of tunneling, bilayer $\nu=1/2$ states must acquire a
magnetization and will not be exactly of the (331) form.
{}From eq.(\ref{energy}) , one can see that the evolution of the (331) state as
the system is tuned from regime ${\bf \{2\}}$ to regime ${\bf \{1\}}$ (either
through increasing layer tunnelling $\Delta$ or reducing the Coulomb
interaction) must lie within the family
\begin{equation}
{\bf d}(s)= \hat{\bf x}_{3}{\rm cos}(\pi s/4) -
i \hat{\bf n}(s) {\rm sin}(\pi s/4) ,   \,\,\,\,\,\, 0\leq s \leq 1,
\label{path}
\end{equation}
where $\hat{\bf n}(s)$ is a real unit vector in the $x_{2}$-$x_{3}$ plane such
that $\hat{\bf n}(1)=\hat{\bf x}_{2}$. The magnetization of
eq.(\ref{path}) is normal to $\hat{\bf x}_{3}$,  meaning that the electron
densities in both layers
remain identical during this process.

A natural question is what the typical magnitude of ${\bf m}$ in the one- to
two-component crossover region.
Although a precise answer can not be given because the functions
$f$, $g_{i}$ and $\lambda$ in eq.(\ref{energy}) are
difficult to calculate, one can estimate the magnitude of ${\bf m}$ from
the recent studies of Wigner crystal states in bilayer systems\cite{Nara}
which allows the spins to rotate freely to response to the tunneling field and
the Coulomb interaction. These studies show that the Wigner crystal at
$\nu=1/2$
(as well as other fillings) acquire a sizable magnetization
(a substantial fraction of the full magnetization
$\overline{\rho}/2$) in a large region of parameter space which covers the
two- to one-component crossover region.

Although the $\nu=1/2$ states generally acquire a magnetization like
the $\nu=1$ state\cite{Yang}\cite{Ho}, they come about for
very different reasons. The$\nu=1$ state is a true ferromagnet. Its
magnetization is spontaneously generated in order to minimize the
Coulomb interactions.
In contrast, the magnetization of the $\nu=1/2$ state
is generated by layer tunneling and the system is in fact more like a
paramagnet.
Despite these  difference, both states share many common magnetic properties.
Probably the most important one is the spin-charge relation.
Following the derivation of the $\nu=1$ case\cite{Yang},
it is straightforward to show that variations
in ${\bf d}$ (hence ${\bf m}$) will lead to density changes of the form
\begin{equation}
\delta \rho({\bf r}) = -|{\bf m}|{\bf \hat{m}}\cdot
\partial_{x}{\bf \hat{m}}\times\partial_{y}{\bf \hat{m}} ,
\label{spincharge} \end{equation}
which is identical to the spin-charge relation of the $\nu=1$ state
(in which case  $|{\bf m}|=(1/2)(1/2\pi)=1/(4\pi)$).
We shall study the consequence of eq.(\ref{spincharge}) on thermodynamic
and transport properties elsewhere, and continue to discuss other symmetry
aspects and implications of the triplet pairing states.

To conclude our symmetry discussions,   we show that the triplet pairing
family possess off diagonal long range order (ODLRO) similar
(but not identical)
to those of single component quantum Hall fluids. In the single component case,
the establishment of ODLRO  is a cruical step in
constructing a Ginzburg-Landau theory\cite{ReadGL}\cite{ZHK}.
The ODLRO of eq.(\ref{tri}) can be extracted in a way similar
to the single-component
case\cite{ReadGL}. We first note that
\begin{equation}
<\rho(z')\rho(z)>_{_{N+1}}=
4(N+1)^{2}\chi^{\ast}_{\alpha\beta}\chi_{\mu\nu}
<\xi_{\beta}(z')\psi_{\alpha}(z')
\psi^{+}_{\mu}(z)\xi^{+}_{\nu}(z)>_{_{N}},
\end{equation}
where $\rho(z)=\psi^{+}_{\mu}(z)\psi^{}_{\mu}(z)$ is the density operator,
$<\rho(z')\rho(z)>_{_{N+1}}$$\equiv$$<N+1,\chi|\rho(z')\rho(z)|N+1,\chi>$, and
$\xi_{\nu}(z)\equiv U^{2}(z)\int (z-a)^{-1}\phi^{+}_{\nu}(a) d^{2}a$,
($[\psi^{+}_{\mu}(z),Q]=2\chi_{\mu\nu}\xi_{\nu}(z)$).
Defining the operator $\overline{\xi}_{\mu}^{+}(z)$ which operates on any
state $|\Psi>$ as
\begin{equation}
\overline{\xi}_{\mu}^{+}(z)|\Psi> \equiv
\frac{\xi^{+}_{\mu}(z)|\Psi>}{<\Psi|[\psi^{+}_{\nu}(z),Q]^{+}
[\psi^{+}_{\nu}(z),Q]|\Psi>^{1/2}} \,\,\,\, ,
\end{equation}
it is easy to show that the matrix $W(z',z)^{}_{\alpha\beta, \mu\nu} \equiv
<N,\chi|\overline{\xi}^{}_{\beta}(z')\psi^{}_{\alpha}(z')
\psi^{+}_{\mu}(z)\overline{\xi}^{+}_{\nu}(z)|N,\chi>$ satisfies
\begin{equation}
\chi^{\ast}_{\alpha\beta}
W(z',z)^{}_{\alpha\beta, \mu\nu}\chi^{}_{\mu\nu}
 =  \frac{<\rho(z)\rho(z')>_{_{N+1}}}{<\rho(z)>_{_{N+1}}^{1/2}
<\rho(z')>_{_{N+1}}^{1/2}}  \end{equation}
\begin{equation}
 \rightarrow
<\rho(z)>_{_{N+1}}^{1/2}<\rho(z')>_{_{N+1}}^{1/2} =\overline{\rho}
= 1/4\pi, \,\,\,\,\,\,\,
{\rm as} \,\,\,
|z-z'|\rightarrow \infty .
\label{ODLRO} \end{equation}
where $\overline{\rho}\equiv<\rho(z)>_{_{N}}=(4\pi)^{-1}$.
Eq.(\ref{ODLRO}) is a statement of ODLRD. It says that the ``density matrix"
$W(z,z')^{}_{\alpha\beta, \mu\nu}$ becomes a product of two functions
$F(z')_{\alpha\beta}\overline{F}(z)_{\mu\nu}$ whose overlap with $\chi_{\mu
\nu}$ is exactly square root of the density.

We now turn to the single layer 5/2 state\cite{Willett}.
Although it is often mentioned in the literature
that this state is a singlet, it is important to
recognize that the current experimental evidence\cite{Willett} (i.e.
suppression of activation energy as the normal of the sample is tilted away
from the magnetic field)
merely demonstrates the two-component nature of the system and is not a proof
of singlet.
Theoretical studies of the 5/2 state have mainly focused on the energetics,
 and are mostly based on
exact diagonalizations of few electron systems. The situation
has not yet been settled\cite{cJain}.
Lastest experiments\cite{Tsui} have in fact shown
that the system remains a good quantum Hall state for tilting angle $\theta$
as large as
$53.3^{o}$ even though the activation energy (which decreases monotonically
as $\theta$) has been suppressed  by a factor
of four. The persistence of the quantum Hall state as a function of $\theta$
is certainly consistent with the picture of a (331) state evolving continuously
towards a Pfaffian state.  The (331) state in this case is made up of equal
number of electrons in the
first Landau level ($E_{1,\uparrow}$) and ($E_{1,\downarrow}$)  of the
up and down spins, with the lowest Landau levels ($E_{0,\uparrow}$ and
$E_{0,\downarrow}$) of both spins completely filled.
The Pfaffian state will be entirely made up of electrons
in $E_{1,\downarrow}$ on top of two fully filled lowest Landau levels.
{}From our previous discussions, it is clear that
the 5/2 state will generally have a magnetization if it is a triplet pairing
state. It can therefore be distinguished from the singlet states by
magnetic resonance methods.

I thank Dan Tsui and Hon Won Jiang
for discussions of the experimental
situation of the 5/2 state. I thank Jainadra Jain for telling his latest
numerical findings, and X.G. Wen for discussions of topological order.
Part of this work was done during the 1994 HKUST Summer School on Strongly
Interacting electron systems. I would like to
thank the Hong Kong University of
Science and Technology for her generous support.


\begin{references}
\bibitem{ATT}
 Y.W. Suen, L.W. Engel, M.B. Santos, M. Shayegan, D.C.
 Tsui, Phys. Rev. Lett. {\bf 68}, 1379 (1992).
J.P. Eisenstein, G.S. Boebinger, L.N. Pfeiffer, K.W. West,
and Song He, Phys. Rev. Lett. {\bf 68}, 1383 (1992).
\bibitem{Willett} R. Willett, J.P. Eisenstein, H.L. Stormer, D.C. Tsui,
A.C. Gossard, and J.H. English, Phys. Rev. Lett. {\bf 59}, 1776 (1987).
\bibitem{Prin2} Y. W. Suen, H. C. Manoharan, X. Ying, M.B. Santos, and
M. Shayegan, Phys. Rev. Lett. {\bf 72}, 3405 (1994).
\bibitem{SongHe} S. He, S. Das Sarma, and X.C. Xie, Phys. Rev. {\bf
B47}, 4394 (1993).
\bibitem{Halperin} B. Halperin, Helv. Phys. Acta. {\bf 56}, 75 (1983).
\bibitem{ReadMoore} G. Moore and N. Read, Nucl. Phys. {\bf B 360}, 362
(1992).
\bibitem{Wen} X.G. Wen and Q. Niu, Phys. Rev. {\bf 41}, 9377 (1990).
B. Blok and X.G. Wen, Phys. Rev. {\bf B 42} 8133 (1990); {\bf B 43} 8337
(1991). M. Greiter, X.G. Wen, and F. Wilczek, Nucl. Phys. {\bf 374}, 567,
(1992).
\bibitem{Jain} J. Jain, Phys. Rev. Lett. {\bf 63}, 199 (1989); Phys. Rev.
{\bf B40}, 8079, {\bf B 41}, 7653 (1990).
\bibitem{Voll} See, for example, {\em The Superfluid Phases of Helilum
3} by D. Vollhardt and P. Wolfle, Taylor and Francis (1990) for general
background of superfluid Helium 3.
\bibitem{ATT2} S.Q. Murthy, J. P. Eisenstein, G.S. Boebinger, L. N.
Pfeiffer, K.W. West, Phys. Rev. Lett. {\bf 72}, 728 (1994).
\bibitem{WenZee} X.G. Wen and A. Zee, Phys. Rev. Lett. {\bf 69}, 1811
(1992);
X.G. Wen and A. Zee, Phys. Rev. B {\bf 47}, 2265 (1993).
Z.F. Ezawa and A. Iwazaki, Int. J. Mod. Phys. B, {\bf 19}, 3205 (1992);
PHys. Rev. B {\bf 47}, 7295 (1993).
\bibitem{Yang} K. Yang, K. Moon, L. Zhang, A.H. MacDonald, S.M.
Girvin, D. Yoshioka, and S.C. Zhang, Phys. Rev. Lett. {\bf 72}, 732,
(1994). (and to be published).
\bibitem{Ho} T.L. Ho, Phys. Rev. Lett. {\bf 73}, 874 (1994).
\bibitem{cHo} We shall use the notation in T.L. Ho, ibid.
\bibitem{Muir} See p.348 in
T. Muir, {\em A Treatise on the Theory of Determinants},
Dover, New York 1960.
\bibitem{ReadGL} N. Read, Phys. Rev. Lett. {\bf 62}, 86, (1989).
\bibitem{Nara} S. Narasimhan and T.L. Ho, {\em Wigner Crystal Phases in
Bilayer Quantum Hall Systems}, submitted to Phys. Rev. B.
\bibitem{ZHK} S.C. Zhang, T.H. Hansson, S. Kivelson, Phys. Rev. Lett. {\bf 62},
82, (1989).
\bibitem{cJain} T. Chakraborty and Pekka Pielilainen, Phys. Rev. {\bf B 38},
10097 (1988).  F.D. M. Haldane and E.H. Rezayi, Phys. Rev. Lett. {\bf 60}, 956
(1988).  A.H. MacDonald, D. Yoshioka, and S.M. Girvin,  Phys. Rev. {\bf B
39}, 8044 (1989). L.Belkhir and J. Jain, Phys. Rev. Lett. {\bf 70} 643 (1993).
The calculate in the last reference show evidence
of some spin polarization as the number of electron
increases.
\bibitem{Tsui} D. Tsui, (private communication).
\end{references}
\end{document}